\documentclass[useAMS,usenatbib]{mn2e}

\usepackage{graphicx}
\usepackage{amssymb}
\usepackage{natbib}
\usepackage{aas_macros}

\title[Detection of $^{12}$CO in ALESS65.1 at $z = 4.44$]{Detection of molecular gas in an ALMA {[CII]}-identified Submillimetre Galaxy at $z = 4.44$} 
\author[M. Huynh et al.] {M.T.~Huynh,$^1$\thanks{E-mail: minh.huynh@uwa.edu.au}
 A.E. Kimball,$^2$ R.P. Norris,$^2$  Ian Smail,$^3$ K.E. Chow,$^2$ \newauthor  
K.E.K. Coppin,$^4$  B.H.C. Emonts,$^{2,5}$ R.J. Ivison,$^{6,7}$  V. Smol{\v c}i{\'c},$^{6,8,9}$ A.M. Swinbank$^3$ \\
$^{1}$ International Centre for Radio Astronomy Research, M468, University of Western Australia, Crawley, WA 6009, Australia \\
$^{2}$ CSIRO Astronomy and Space Science, PO Box 76, Epping, NSW, 1710, Australia \\
$^{3}$ Institute for Computational Cosmology, Department of Physics, Durham University, Durham DH1 3LE, UK \\
$^{4}$ Centre for Astrophysics, University of Hertfordshire, College Lane, Hatfield, Hertfordshire, UK AL10 9AB \\
$^{5}$ Centro de Astrobiolog{\'i}a (INTA-CSIC), Ctra de Torrej{\'o}n a Ajalvir, km 4, E-28850 Torrej{\'o}n de Ardoz, Madrid, Spain \\
$^{6}$ European Southern Observatory, Karl Schwarzschild Strasse 2, D-85748 Garching, Germany \\
$^{7}$ Institute for Astronomy, University of Edinburgh, Royal Observatory, Blackford Hill, Edinburgh EH9 3HJ, UK \\
$^{8}$ Argelander Institut for Astronomy, Auf dem Hugel 71, Bonn D-53121, Germany \\
$^{9}$ Physics Department, University of Zagreb, Bijeni{\v c}ka cesta 32, 10002 Zagreb, Croatia}

\begin{document}

\maketitle

\label{firstpage}

\begin{abstract}

We present the detection of $^{12}$CO(2--1) in the $z = 4.44$ submillimetre galaxy ALESS65.1 using the Australia Telescope Compact Array. 
A previous ALMA study of submillimetre galaxies in the Extended {\it Chandra} Deep Field South determined the redshift of this optically and near-infrared undetected source through the measurement of [C{\sc II}] 157.74 $\mu$m emission. 
Using the luminosity of the $^{12}$CO(2--1) emission we estimate the gas mass to be $M_{\rm gas} \sim 1.7 \times 10^{10}$ ${\rm M}_\odot$. The gas depletion timescale of ALESS65.1 is $\sim$ 25 Myr, similar to other high redshift submillimetre galaxies and consistent with $z > 4$ SMGs being the progenitors of massive ``red-and-dead" galaxies at $z > 2$. The ratio of the [CII],  $^{12}$CO  and far-infrared luminosities implies a strong far-ultraviolet field of $G_0 \sim 10^{3.25}$,  which is at the high end of the far-ultraviolet fields seen in local starbursts, but weaker than the far-ultraviolet fields of most nearby ULIRGs. The high ratio of $L_{\rm [CII]}/L_{\rm FIR} = 1.0 \times 10^{-3}$ observed in ALESS65.1, combined with $L_{\rm [CII]}/L_{\rm CO} \sim 2300$, is consistent with ALESS65.1 having more extended regions of intense star formation than local ULIRGs.

\end{abstract}


\section{Introduction}

Submillimetre galaxies (SMGs) are a population of ultraluminous infrared galaxies with extreme star formation rates (SFRs)  of 100 -- 1000 $\rm{M}_\odot$ yr$^{-1}$ (e.g. \citealp{blain2002}) with typical redshifts of $z \sim 2.5$ (e.g. \citealp{chapman2005, wardlow2011,yun2012,smolcic2012,simpson2014}).  
About 20\% of SMGs lie at $z > 4$ \citep{wardlow2011,smolcic2012,simpson2014}, and a few dozen of these have now been studied in detail  \citep{capak2008,capak2011,daddi2009b,daddi2009a,coppin2009,knudsen2010, carilli2010, carilli2011,riechers2010,smolcic2011,cox2011,combes2012, walter2012,weiss2013, vieira2013}. 
These sources are interesting as they represent the earliest examples of extreme starburst events in massive galaxies, and knowledge of their star formation activity and gas content is crucial for understanding the growth of massive ellipticals. 

Radio emission produced by the rotational transition of carbon monoxide ($^{12}$CO) is one of the most accessible tracers of cold molecular gas in galaxies \citep{carilli2013}.
 Detections of $^{12}$CO in $z > 4$ SMGs have shown they are gas rich systems with sufficient reservoirs ($M_{\rm gas} > 10^{10}$ $\rm{M}_\odot$) to sustain the extreme star formation rates of $\sim$ 1000 $\rm{M}_\odot$ yr$^{-1}$ for only short time scales (10s of Myrs) \citep{coppin2010, riechers2010, huynh2013}, unless the gas is replenished. High redshift SMGs are therefore seen as the likely progenitors of the luminous red galaxies seen at $z > 2$ \citep{cimatti2008}.

A powerful emission line for studying the ISM of high redshift sources is the $^2P_{3/2}$ -- $^2P_{1/2}$ fine structure line of singly ionised carbon at 157.74 $\mu$m (hereafter [CII]), which can represent up to 1\% of the bolometric luminosity of star forming galaxies (e.g. \citealp{crawford1985, stacey1991}). 
This line emission arises predominately from the edges of molecular clouds illuminated by the UV photons of young-massive stars (i.e. photodissociation regions), but a non-negligible contribution can also come from HII regions and the more diffuse warm interstellar medium \citep{madden93,heiles94}. The [CII] line therefore provides an important probe of the physical conditions in the ISM of a galaxy. This carbon line has now been studied in several high redshift SMGs \citep{debreuck2011,wagg2012,walter2012,rawle2014,debreuck2014}. 

An ALMA Cycle 0 study of 126 submillimetre sources located in the LABOCA Extended {\it Chandra} Deep Field South (LESS, \citealp{weiss2009, karim2013, hodge2013}) resulted in the serendipitous identification of [CII] line emission from two SMGs (\citealp{swinbank2012}, hereafter S12). The high [CII]/far-infrared luminosity ratio of these two SMGs, roughly ten times higher than that observed in local ultraluminous infrared galaxies, was interpreted as evidence that their gas reservoirs are more extended (S12).  
High [CII]/far-infrared ratios (\citealp{stacey2010, ivison2010b}; S12) add to the mounting evidence that star-formation in SMGs takes place in a region larger than the compact nuclear starbursts of local ULIRGs; which includes extended radio morphologies (e.g. \citealp{chapman2004,biggs2008}), extended H-$\alpha$ morphologies (e.g. \citealp{swinbank2006}), and large $^{12}$CO(1--0) sizes (e.g. \citealp{ivison2010a, hodge2012}). 

In \cite{huynh2013} we presented Australia Telescope Compact Array (ATCA) observations, totalling about 20 hours on-source, of one of the ALMA detected SMGs, ALESS J033252.26-273526.3 (hereafter ALESS65.1). No $^{12}$CO(2--1) emission was detected but we were able to place constraints on the gas mass and physical conditions of the gas using the ALMA [CII] detection. Further observations were obtained with the ATCA in 2013 and, combined with the previous data, we now have detection of $^{12}$CO(2--1). This paper presents the ATCA observations, data analysis, and a discussion of the physical parameters derived from the molecular gas detection. We adopt the standard $\Lambda$-CDM cosmological parameters of $\Omega_{\rm M} = 0.27$,  $\Omega_{\rm \Lambda} = 0.73$, and a Hubble constant of 71 km s$^{-1}$ Mpc$^{-1}$ throughout this paper.

\section{Observations and Results}

The $^{12}$CO(2--1) line ($\nu_{\rm rest}$ = 230.538 GHz) in ALESS65.1 (RA = 03 32 52.26, Dec = $-$27 35 26.3, J2000) (S12) was observed over a period of four nights in August 2012 and three nights in July 2013 with the Australia Telescope Compact Array (ATCA), using the Compact Array Broadband Backend (CABB). During both runs the array was in the most compact five-antenna configuration, H75, which has a maximum baseline of 89m and two antennas set along a northern spur. This hybrid configuration allows good $(u,v)$ coverage to be obtained for integrations less than the full 12 hour synthesis.  The 7mm receiver was centered on 42.343 GHz, the expected frequency of the $^{12}$CO(2--1) line emission given the [CII] redshift of $z = 4.4445$ (S12). The 2GHz bandwidth of CABB results in a frequency coverage of approximately 41.3 to 43.3 GHz, covering $^{12}$CO(2--1) emission between $z$ = 4.32 -- 4.58.  The weather was average to good, with rms atmospheric path length variations of 100 to 400 microns throughout the runs, as measured on the 230m baseline ATCA Seeing Monitor \citep{middelberg2006}.
Following \cite{emonts2011}, a bandpass calibration scan was acquired at the beginning and end of each 8 hour night. Phase and amplitude calibration information was acquired with 2 minute scans on PKS 0346$-$279 every 10 minutes and pointing checks performed on the same source every hour. For flux calibration we observed Uranus at the beginning of the nights, at an elevation of $\sim$55 degrees. The uncertainty in the flux density calibration using the standard {\sc miriad} model of Uranus is estimated to be 30\% \citep{emonts2011}. 

The data were calibrated, mapped and analysed using the standard {\sc miriad} \citep{sault1999} and {\sc karma} \citep{gooch1996} packages.  The synthesized beam from natural weighting is 14.0 $\times$ 9.0 arcsec. A total of about 40 hours on-source integration time was obtained over the 7 $\times$ 8 hour nights. 
ALESS65.1 was not detected in the 42.3 GHz continuum map from the full CABB band, which achieves an rms noise level of 7.4 $\mu$Jy beam$^{-1}$. 

The resultant channel noise in the 1 MHz (7.1 km s$^{-1}$) wide spectrum is $\sim$ 0.29 mJy beam$^{-1}$, consistent with other comparable 7mm ATCA/CABB surveys (e.g. \citealp{coppin2010, emonts2014}) and the ATCA online sensitivity calculator.  The visibilities were resampled to velocity resolutions of 200, 400 and 600 km s$^{-1}$ and each cube was examined for an emission line near the ALMA position. 
The spectra at the source position in the 200, 400 and 600 km~s$^{-1}$ binned cubes (Figure 1) have an rms of  0.057, 0.042 and 0.031 mJy beam$^{-1}$, respectively. 
We identify a line at the ALMA position and redshift in all 3 cubes. The line is detected at more than the 3$\sigma$ level across multiple channels in the 200 and 400 km~s$^{-1}$ cubes and at the 5.1 $\sigma$ level in the 600 km s$^{-1}$ cube (see Figure 2). From a Gaussian fit to the line we find an integrated line flux density of 112 $\pm$ 35 mJy km s$^{-1}$, a FWHM of 620 $\pm$ 120 km s$^{-1}$, and a small offset from the [CII] of 140 $\pm$ 70 km s$^{-1}$.

\begin{figure*}
\centering
\includegraphics[width=0.8\columnwidth]{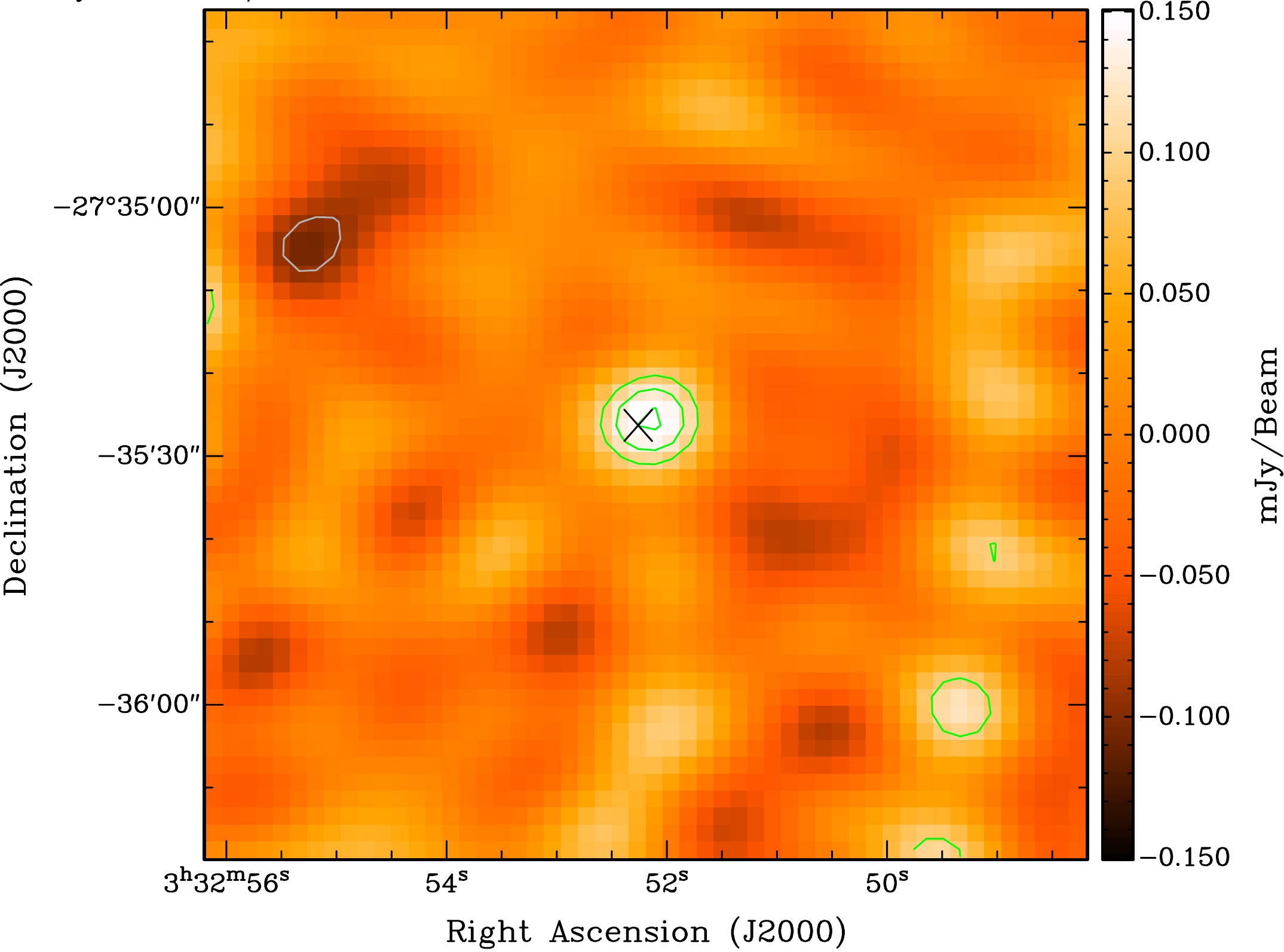}
\hspace{10mm}
\includegraphics[width=0.8\columnwidth]{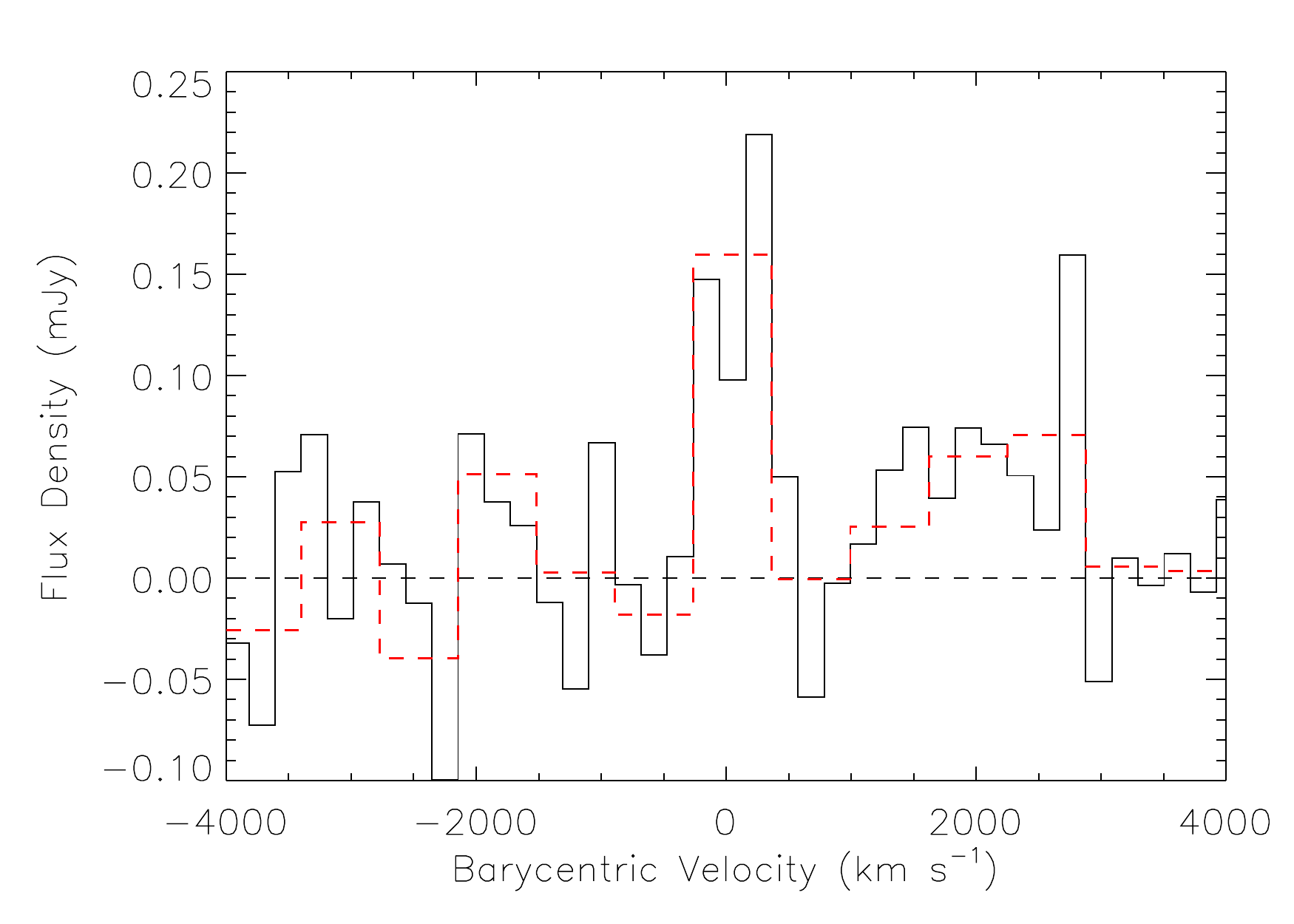}
\caption{Left: $^{12}$CO(2--1) emission line map of ALESS65.1, using 600 km s$^{-1}$ binning to get all the flux in the one channel. The colour scale is -0.150 mJy (black) to 0.150 mJy (white). The green contours are 3, 4 and 5$\sigma$, and the grey contour is $-$3$\sigma$. The cross marks the position of the ALMA source. We clearly detect $^{12}$CO(2--1) emission associated with the ALMA source. Right: The $^{12}$CO(2--1) spectrum of ALESS65.1 binned into 200 km s$^{-1}$ channels and extracted at the ALMA position (black solid line). The red dashed line shows the spectrum binned into 600 km s$^{-1}$ channels for maximum sensitivity.}
\label{fig:spectra}
\end{figure*}

\section{Discussion}

The observed and derived properties of ALESS65.1 are summarised in Table 1. The total cold gas (H$_2$ + He) mass in ALESS65.1 can be estimated from the CO(2--1) line luminosity.  Following \cite{solomon2005}, the line luminosity is $L'_{\rm CO(2-1)}$ = 2.16 $\pm$ 0.67 $\times$ $10^{10}$ K km s$^{-1}$ pc$^2$. If we assume the gas is thermalised (i.e. intrinsic brightness temperature and line luminosities are independent of $J$ transition), so $L'_{\rm CO(2-1)}$ = $L'_{\rm CO(1-0)}$, and a CO-to-H$_2$ conversion factor $\alpha = 0.8$ ${\rm M}_\odot$ (K km s$^{-1}$ pc$^2$)$^{-1}$, the total cold gas mass is estimated to be $M_{\rm gas}$ = 1.7 $\pm$ $0.5 \times 10^{10}$ ${\rm M}_\odot$. This is consistent with the gas mass found in other $z > 4$ SMGs \citep{schinnerer2008, daddi2009b, coppin2010, walter2012}. We caution that this gas mass is dependent on the poorly-known CO-to-H$_2$ conversion factor $\alpha$. Discs like the Milky Way have relatively high values of $\alpha$ $\sim$ 3 -- 5, while a smaller value,  $\alpha = 0.8$, is appropriate for local ULIRGs (e.g. \citealp{downes1998}) and is the value commonly adopted for high redshift SMGs (but \citealp{bothwell2013} assume $\alpha = 1$). 

The gas and stellar mass estimates for ALESS65.1 can be combined to derive a total baryonic mass of the galaxy. The stellar mass of the galaxy was estimated from the rest-frame absolute $H$-band magnitude to be $M_* \sim 9 \times 10^{10}$ ${\rm M}_\odot$ (S12), so the gas fraction is modest with $M_{\rm gas}/M_* \sim 0.2$. The total baryonic mass $M_{\rm bary}$ = $M_{\rm gas} + M_*$ is $\sim$$10.7 \times 10^{10}$ ${\rm M}_\odot$. This is consistent with the dynamical mass for ALESS65.1, based on the spatial extent of the marginally resolved [CII] line, of $M_{\rm dyn} \sin^2(i) \sim (3.4 \pm 1.8) \times 10^{10}$ ${\rm M}_\odot$ (S12). The baryonic and dynamical mass estimates suggest an inclination angle of about 30 degrees for ALESS65.1, albeit with considerable uncertainty.   

ALESS65.1 is detected at 870$\mu$m in continuum by ALMA and marginally detected by {\it Herschel} (S12). Its total restframe IR (8 -- 1000 $\mu$m) luminosity was estimated in S12 to be $(2.0\pm0.4) \times 10^{12}$ ${\rm L}_\odot$. A more careful deblending analysis has determined the source is not detected in {\it Herschel} photometry \citep{swinbank2014}. We estimate a revised IR luminosity for ALESS65.1 by fitting the IR SED using the method of \cite{swinbank2014} but with a fixed redshift of $z = 4.44$, resulting in LIR (8 -- 1000 $\mu$m) = $(3.9^{+1.8}_{-1.5}) \times 10^{12}$ ${\rm L}_\odot$ and LIR (42 -- 122 $\mu$m) = $(3.1^{+1.9}_{-1.6}) \times 10^{12}$ ${\rm L}_\odot$. This is greater than the IR luminosity estimate of S12 due to the much better modelling  
of the detection limits of the  {\it Herschel} SPIRE maps. This IR luminosity corresponds to a star formation rate of $\sim$ $(670 \pm 310)$ ${\rm M}_\odot$ yr$^{-1}$ using the conversion of \cite{kennicutt1998}. The gas depletion timescale, $\tau = M_{\rm gas} /{\rm SFR} = 25$ $\pm$ 15 Myr, is similar to the gas depletion rates of other high redshift SMGs \citep{schinnerer2008,coppin2010}. 
Assuming there is no further gas infall and there is 100\% efficiency in converting the gas to stars, the star formation is effectively shut off at $z \sim 4.4$ and this galaxy would appear red and dead by $z \sim 3$ (750 Myr after gas depletion). This short gas depletion timescale therefore provides further evidence that $z > 4$ SMGs have the gas consumption timescales necessary to be the progenitors of the most distant red-and-dead ``ellipticals", those found at $z \gtrsim 3$ \citep{marchesini2010,muzzin2013}. 
  
We next examine the physical conditions of the gas in ALESS65.1 using the [CII]  and $^{12}$CO(2--1) detections. The  $L_{\rm [CII]}/L_{\rm FIR}$  versus $L_{\rm CO(1-0)}/L_{\rm FIR}$ diagram is a powerful diagnostic as these two ratios are sensitive to gas density $n$ and the incident far-ultraviolet (FUV) flux $G_0$ \citep{stacey1991}. Figure 2 shows ALESS65.1 compared with other low and high redshift galaxies, and solar metallicity photodissociation region (PDR) model curves \citep{kaufman1999}. This diagram can be used to roughly estimate both $n$ and $G_0$ for a galaxy, but with the following assumptions: (i)  the [CII] emission comes mainly from PDRs, with little contribution from the diffuse ionised medium or cosmic-ray-heated gas, and (ii) AGN and their related X-Ray Dissociation Regions (XDRs) do not contribute significantly to the FIR and [CII] luminosity. To be consistent with both \cite{debreuck2011} and \cite{stacey2010} in Figure 2 we assume $L_{\rm CO(2-1)} / L_{\rm CO(1-0)} = 7.2$, which is 90\% of its value if the gas was fully thermalised and optically thick. We note that this is consistent with \cite{bothwell2013} who find  $L_{\rm CO(2-1)} / L_{\rm CO(1-0)} = 6.72 \pm  1.04$ for $z \sim 2$ SMGs. Cosmic ray rates are greater in starbursts compared to normal galaxies but this does not seem to result in higher [CII]/CO ratios, so cosmic ray ionization does not appear to dominate the [CII]/CO ratio in local galaxies \citep{debreuck2011}. In \cite{huynh2013} we found that ALESS65.1 is not detected in the 250ks {\it Chandra} X-Ray observations of this region \citep{lehmer2005}, so it is not an unobscured luminous QSO ($L_{\rm 3-44 keV} \lesssim 2$--3$ \times 10^{44}$ erg s$^{-1}$, for $N_{\rm H} = 0$ -- 10$^{23.5}$ cm$^{-2}$) (see also \citealp{wang2013}).
Furthermore, a decomposition of the mid-infrared and far-infrared SED implied that the AGN contribution to the total FIR luminosity is  $\lesssim$10\% \citep{huynh2013}. ALESS65.1 therefore appears to be dominated by star-formation processes and the AGN contribution to [CII] and  $L_{\rm FIR}$ is likely to be minimal. 

\begin{table}
\centering
\caption{Observed and derived properties of ALESS65.1}
\begin{tabular}{lcc}  \hline
Parameter & Value & Reference \\ \hline
$z_{\rm [CII]}$ & 4.4445 $\pm$ 0.0005 & S12\\
$I_{\rm [CII]}$ & 5.4 $\pm$ 0.7 Jy km s$^{-1}$ & S12\\  
FWHM$_{\rm [CII]}$ & 470 $\pm$ 35 km s$^{-1}$ & S12\\
$L_{\rm [CII]}$ & (3.2 $\pm$ 0.4) $\times$ $10^9$ ${\rm L}_\odot$ & S12 \\
$L_{\rm FIR}$& (3.1$^{+1.9}_{-1.6}$) $\times$ $10^{12}$ ${\rm L}_\odot$ & this paper \\
$I_{\rm CO(2-1)}$ & $ 0.112 \pm 0.035$ Jy km s$^{-1}$ & this paper \\  
FWHM$_{\rm CO(2-1)}$ & 620 $\pm$ 120 km s$^{-1}$ & this paper\\
$M_{\rm gas}$ & (1.7 $\pm$ 0.5) $\times$ $10^{10}$ ${\rm M}_\odot$ & this paper \\
$L_{\rm CO(2-1)}$ &  (8.45 $\pm$ 2.64) $\times$ $10^{6}$ ${\rm L}_\odot$  & this paper \\
$L'_{\rm CO(2-1)}$ &  (2.16 $\pm$ 0.67) $\times$ $10^{10}$ K km s$^{-1}$ pc$^2$ & this paper \\
\hline
\end{tabular}
\end{table}

In examining the PDR physical conditions, we multiply the $^{12}$CO(2--1) flux by a factor of two to account for detecting CO emission only from the illuminated PDR side \citep{kaufman1999, hailey-dunsheath2010}, and also multiply the [CII] flux by a factor of 0.7 to remove non-PDR contributions (e.g. \citealp{hailey-dunsheath2010, stacey2010}). The $^{12}$CO geometry correction applies to all galaxies in Figure 2, and so does not affect the relative position of ALESS65.1 on the diagram compared to other galaxies. 
Using the \cite{kaufman1999} models,  we find ALESS65.1 has $G_0 \sim 10^{3.25}$ (where $G_0$ is in units of the Habing Field, $1.6 \times 10^{-3}$ ergs cm$^{-2}$ s$^{-1}$) and $n \sim10^{4.6}$ cm$^{-3}$ (Figure 2). Such a FUV radiation field is higher than the FUV fields seen in low redshift normal galaxies, but it is consistent with the strong FUV fields seen some local starbursts and $z > 1$ galaxies. 
The $z > 4$ SMGs shown in Figure 2 have inferred FUV fields at the low-end of nearby ULIRGs, the local analogues of SMGs. 
The estimates of $G_0$ and $n$ implies a PDR temperature $\sim$300 K \citep{kaufman1999} for ALESS65.1. Using Equation 1 from \cite{hailey-dunsheath2010}, we estimate the atomic gas associated with the PDR to be approximately 3 $\times$ $10^9$ $M_{\odot}$, which is $\sim$ 20\% of the total cold gas mass. This is similar to the fraction found in a similar SMG, LESS J033229.4, at $z \sim 4.76$ \citep{debreuck2014}, redshift 1 -- 2  starforming galaxies \citep{stacey2010} and local IR bright galaxies \citep{stacey1991}. 
  
\begin{figure}
\includegraphics[width=0.49\textwidth]{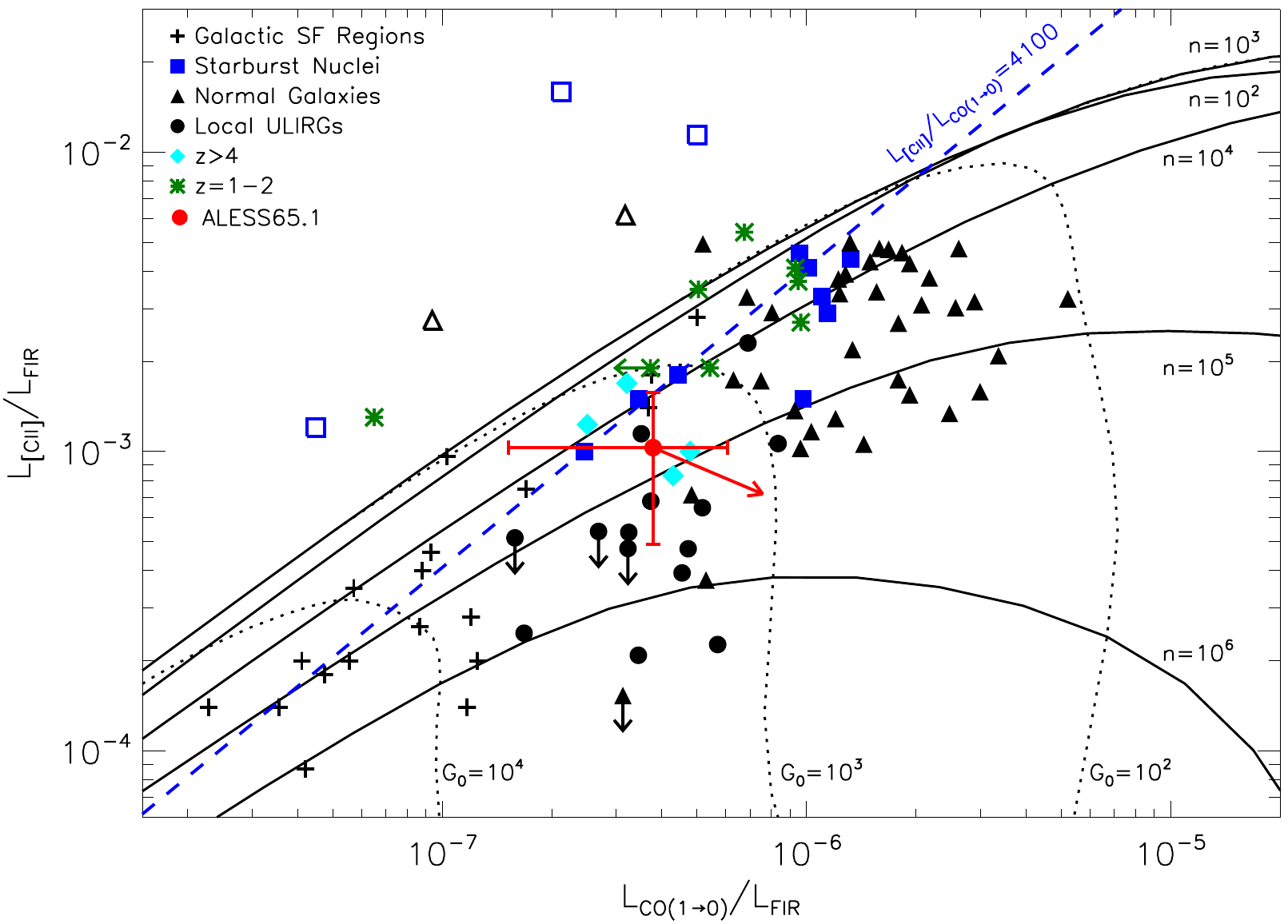}
\caption{$L_{\rm [CII]}/L_{\rm FIR}$  versus $L_{\rm CO(1-0)}/L_{\rm FIR}$ for ALESS65.1 (red point with upper limit on $L_{\rm CO(1-0)}/L_{\rm FIR}$) compared with Galactic star forming regions, starburst nuclei, normal galaxies, local ULIRGs, and high redshift ($z > 1$) sources. Empty symbols indicate low metallicity sources, which lie at high $L_{\rm [CII]}/L_{\rm CO(1-0)}$. Black lines represent the solar metallicity PDR model calculations for gas density ($n$) and FUV field strength ($G_0$) from Kaufman et al. (1999). This figure is adapted from Stacey et al. (2010) with additional data from Walter et al. (2012), Wagg et al. (2012, 2014), De Breuck et al. (2014) and Rawle et al. (2014). ALESS65.1 has a higher $L_{\rm [CII]}/L_{\rm FIR}$ ratio than that found in most local ULIRGs, consistent with other $z > 4$ SMGs, but its $L_{\rm [CII]}/L_{\rm CO}$ ratio is similar to local starbursts and other $z > 1$ sources. The red arrow indicates geometric corrections to the CO emission and non-PDR corrections to the [CII] emission, but this would apply to all galaxies in the figure and so does not affect the relative position of ALESS65.1 on the diagram compared to the other galaxies.}
\label{fig:cii_co}
\end{figure} 
 
Local starbursts and Galactic OB star forming regions lie on a line with [CII]/CO luminosity ratios of about 4100 in Figure 2. Higher [CII]/CO ratios can be found in low metallicity systems, such as 30 Doradus in the LMC, where the size of the [CII] emitting envelope of the cloud (relative to the the CO emitting core) is much larger than in more metal-rich systems \citep{stacey1991}. 
Metallicity is expected to affect the [CII]/CO ratios of the highest redshift galaxies because the interstellar mediums of the youngest galaxies are expected to be less enriched by supernova. 
However such enrichment appears to occur very quickly. For example, LESS J033229.4 at $z = 4.76$ has a [CII]/CO ratio of $\sim$ 5000 (De Breuck et al. 2014), suggesting near solar metallicity (see also \citealp{nagao2012}).  ALESS65.1 has  $L_{\rm [CII]}/L_{\rm CO} \sim 2700$, which also indicates that the gas is not of low metallicity. 
  
\begin{figure}
\includegraphics[width=0.49\textwidth]{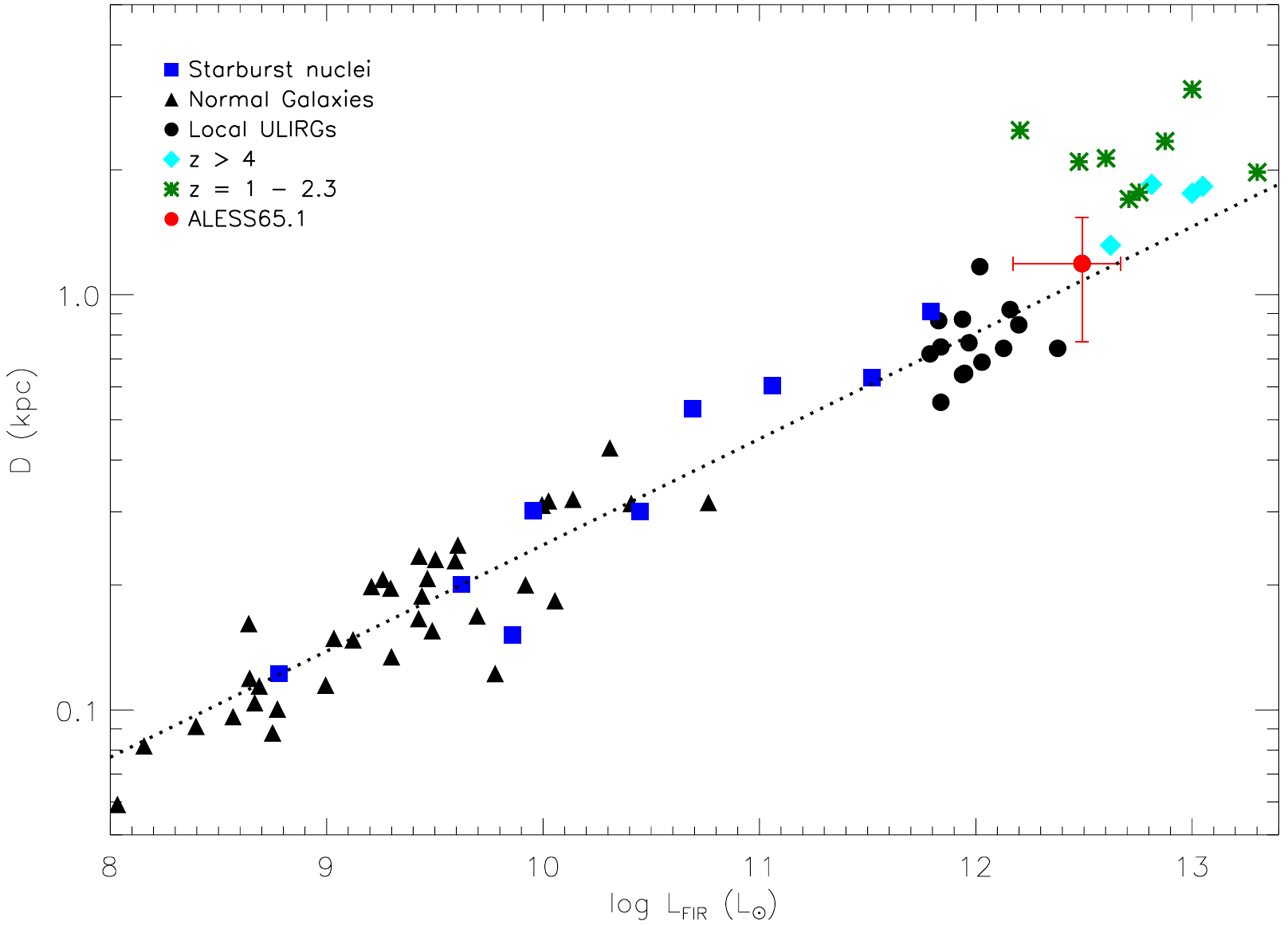}
\caption{Size, D, versus $L_{\rm FIR}$ for galaxies in Figure \ref{fig:cii_co}. The size is estimated using $G_0\propto \lambda L_{\rm FIR}/D^3$ and M82 physical parameters. $G_0$ is estimated from the [CII]/$L_{\rm FIR}$ and CO(1--0)/$L_{\rm FIR}$ ratios and Kaufman et al. (1999) PDR models (Figure \ref{fig:cii_co}). Galaxies at $z > 1$ appear to lie above the local ``trend" (dashed line), which supports the hypothesis that they have more extended starforming regions.}
\label{fig:size}
\end{figure}

The FUV radiation fields, $G_0$, of ALESS65.1 and four other $z > 4$ SMGs shown in Figure 2, LESS J033229.4 (De Breuck et al. 2014), HDF850.1 (Walter et al. 2012), BRI 1202-0725 \citep{wagg2012,wagg2014}, and HLS0918 \citep{rawle2014}, are similar to that of local starbursts, but these distant galaxies have a much higher FIR luminosity, leading to suggestions that they are scaled-up versions of local starbursts. For a given $L_{\rm FIR}$ the size of the emission region will increase for smaller $G_0$. Following Stacey et al. (2010), we scale up from M82 using two laws from \cite{wolfire1990} to constrain the size: $G_0 \propto \lambda L_{\rm FIR}/D^3$ if the mean free path of a UV photon $\lambda$ is small and  $G_0 \propto L_{\rm FIR}/D^2$ if the mean free path of a UV photon is large. Applying these relations and using $G_0 = 10^{3.25}$ for ALESS65.1 yields a diameter of 1.2 -- 2.4 kpc. This is consistent with the marginally resolved [CII] data which shows ALESS65.1 has a possible extent of 3.3 $\pm$ 1.7 kpc (S12). 
We use the $G_0 \propto \lambda L_{\rm FIR}/D^3$ relation and M82 scaling parameters to estimate sizes of the starforming regions of galaxies plotted in Figure 2. 
Figure 3 shows that the starburst in all $z > 4$ SMGs appears to be extended over galactic scales, and galaxies at $z > 1$ appear to lie above the locus delineated by local normal galaxies, local starbursts and nearby ULIRGs (Figure \ref{fig:size}), with moderate redshift ($1 < z < 2.3$) SMGs lying more above the local `trend' than the $z > 4$ SMGs. This further suggests that starburst regions in distant galaxies are larger than in local galaxies of similar luminosity. There are caveats in interpreting Figure \ref{fig:size} however: the $G_0$ estimate has significant uncertainties because of the large uncertainties in the line luminosity ratios used in Figure 2; the scaling constant maybe different to that of M82 for the different galaxy samples; and the sample sizes at $z > 1$ and $z > 4$ are small and have no galaxies with $ L_{\rm FIR} < 10^{12}$ ${\rm L}_\odot$. 

In this work we have assumed a single-phase ISM, but ALESS65.1 may have a multi-phase ISM. The spatially-resolved $^{12}$CO,  $^{13}$CO,  and C$^{18}$O  study of the gravitationally-magnified SMG SMM\,J2135$-$0102 found that it has an ISM best described by a warm compact component surrounded by a cooler more extended one \citep{swinbank2011,danielson2011,danielson2013}. A multi-component ISM analysis is not currently possible due to the limitations of existing $^{12}$CO data, but future higher-resolution line studies with ALMA may find a similar multi-component ISM for ALESS65.1.

\section{Conclusion}

We have observed ALESS65.1 for 40 hours to search for $^{12}$CO(2--1) emission in this $z = 4.44$ submillimetre galaxy. 
The line is detected at a $\sim$ 5$\sigma$ level with an integrated line flux density of 112 $\pm$ 35 mJy km s$^{-1}$ and a FWHM of 620 $\pm$ 120 km s$^{-1}$. 
We find a $^{12}$CO(2--1) line luminosity of $L_{\rm CO(2-1)} = 8.5 \times 10^{6}$ ${\rm L}_\odot$ and a cold gas mass of $M_{\rm gas} \sim 1.7 \times 10^{10}$ ${\rm M}_\odot$.  This implies a gas depletion timescale in ALESS65.1 of $25$ Myr, comparable to other $z > 4$ SMGs and consistent with this high redshift population being the progenitors of $z \sim 3$ red-and-dead galaxies.

We examined the physical conditions of the gas in ALESS65.1 using the $L_{\rm [CII]}/L_{\rm FIR}$  versus $L_{\rm CO(1-0)}/L_{\rm FIR}$ diagram. We find ALESS65.1 has a strong FUV field $G_0 \sim 10^{3.25}$ comparable to some local starbursts, but lower than that seen in most nearby ULIRGs, the local population with IR luminosities similar to ALESS65.1. The observed [CII] to FIR ratio, $L_{\rm [CII]}/L_{\rm FIR} = 1.0 \times 10^{-3}$, is high compared to local ULIRGs (as noted by S12). Combined with $L_{\rm [CII]}/L_{\rm CO} \sim 2700$, this high [CII] to FIR ratio is consistent with ALESS65.1 having more extended regions of intense star-formation than local ULIRGs. The [CII]/CO ratio provides no evidence for low metallicity gas in ALESS65.1. 

A larger study of [CII] and $^{12}$CO in distant starbursts is needed to confirmed whether the majority of $z > 4$ starbursts have enhanced [CII] emission compared to their local analogues, and whether this is because of metallicity effects, the relative size of PDR regions, or other effects.  Future ALMA surveys will shed further light on the physical conditions of the gas in star forming galaxies in the early universe.  

\section*{Acknowledgments}

IRS acknowledges support from the STFC (ST/I001573/1), the ERC Advanced Investigator Programme DUSTYGAL 321334 and a Royal SocietyWolfson Merit Award. VS acknowledges the Group of Eight (Go8) fellowship and funding from the European Union's Seventh Frame-work program under grant agreement 337595 (ERC Starting Grant, 'CoSMass'). BE acknowledges funding through MINECO grant AYA2010-21161-C02-01. The Australia Telescope is funded by the Commonwealth of Australia for operation as a National Facility managed by CSIRO.

\bibliographystyle{mn2e}
\bibliography{refs}

\bsp

\label{lastpage}

\end{document}